\documentclass[fleqn,twoside]{article}
\usepackage[headings]{espcrc2}

\readRCS
$Id: espcrc2.tex,v 1.2 2004/02/24 11:22:11 spepping Exp $
\usepackage{graphicx}
\usepackage[figuresright]{rotating}

\def\ve{\varepsilon}

\newcommand{\AmS}{{\protect\the\textfont2
  A\kern-.1667em\lower.5ex\hbox{M}\kern-.125emS}}

\title{
Expansion of massive scalar one-loop integrals to 
${\cal{O}}(\varepsilon^{2})$}

\author{J\"urgen G. K\"orner\address[MCSD]{Institut f\"ur Physik,  Johannes Gutenberg-Universit\"at, D55099 - Mainz, Germany},
        Zakaria Merebashvili\address{High Energy Physics Institute,
Tbilisi State University, 380086 Tbilisi, Georgia}\thanks{Supported in part by the
        Deutsche Forschungsgemeinschaft DFG under contract 436 GEO 17/3/03
        and 436 GEO 17/4/04.},
        and
        M. Rogal\addressmark[MCSD]\thanks{Supported by the Graduiertenkolleg
        ``Eichtheorien'' at the University of Mainz.}}
       
\begin{document}

\begin{abstract}
We report on the results of an ongoing calculation of massive scalar one-loop
one-, two-, three- and four-point integrals up to ${\cal{O}}(\varepsilon^{2})$
which are needed in the NNLO calculation of heavy hadron production.
\vspace{1pc}
\end{abstract}

\maketitle

\section{Introduction}
The full next-to-leading order (NLO) QCD corrections to hadroproduction of heavy 
flavors have been completed as early as 1988 \cite{Dawson:1988,been}. They 
have raised the leading order (LO) estimates \cite{LO:1978} but were 
still below  the experimental results (see e.g. \cite{Italians}).
In a recent analysis theory moved closer to experiment \cite{Italians}. 
A large uncertainty in the NLO calculation results from the freedom in the 
choice of the renormalization and factorization scales.
The dependence on the factorization and renormalization scales is expected to be 
greatly reduced at next-to-next-to-leading order (NNLO). This will
reduce the theoretical uncertainty. Furthermore, one may hope that  there is yet  
better agreement between theory and experiment at NNLO.
\section{General remarks}
\begin{figure}[t]
\center
\includegraphics[height=9cm]{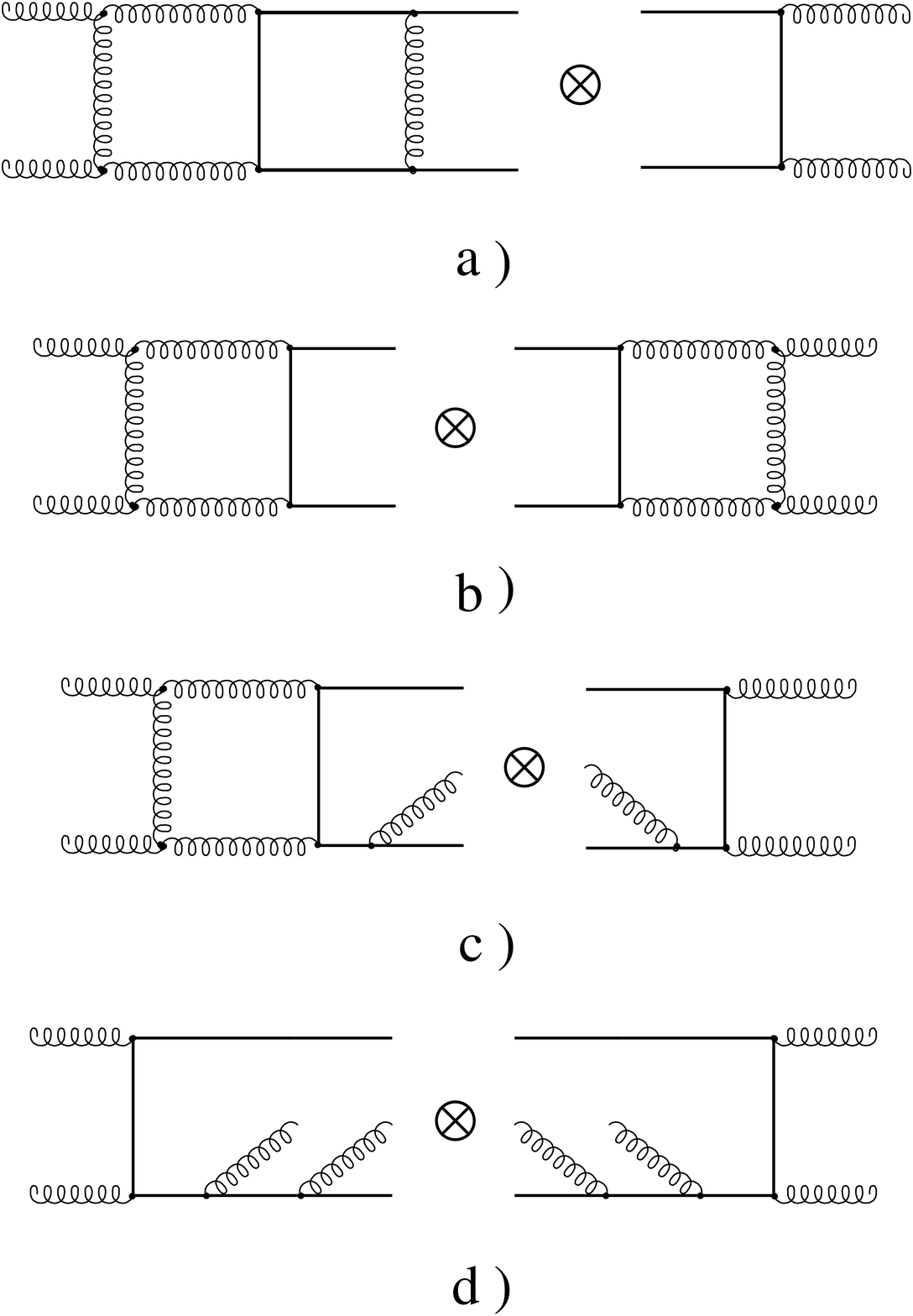}
\caption{Exemplary gluon fusion diagrams for the  NNLO calculation of heavy 
hadron production}
\label{nnlo}       
\end{figure}

In Fig.~\ref{nnlo} we show one generic diagram each for the four classes
of gluon induced contributions that need to be calculated for the NNLO corrections
to hadroproduction of heavy flavors. They involve the two-loop
contribution (Fig.~\ref{nnlo}a), the loop-by-loop contribution (Fig.~\ref{nnlo}b), 
the one-loop gluon emission contribution (Fig.~\ref{nnlo}c) and, finally, the two 
gluon emission contribution (Fig.~\ref{nnlo}d). We mention that there is an 
interesting subclass of the diagrams 
in Fig.~\ref{nnlo}c where  the outgoing gluon is attached 
directly to the loop. One then has a five-point function which, when folded with the
corresponding tree graph contribution, has to be 
calculated up to ${\cal O}(\ve^{2})$.

In our work we have concentrated on the
loop-by-loop contributions exemplified by Fig.~\ref{nnlo}b. Specifically, working 
in the framework of  dimensional regularization, we are in the process of 
calculating ${\cal O}(\ve^2)$ results for
all scalar massive one-loop one-, two-, three- and four-point integrals that are
needed in the calculation of hadronic heavy flavour production. It should be clear 
from the flavour flow of the diagrams that, apart from 
the two mass scales set by the kinematics of the process there is only one explicit heavy mass scale in the problem.  
The integration is generally done by writing down the Feynman parameter 
representation 
for the corresponding integrals, integrating over Feynman parameters up to 
the last remaining integral, expanding the integrand of the last remaining 
parametric integral in terms of the dimensional parameter $\ve$ and doing the 
last parametric integration on 
the coefficients of the expansion.
Because the one-loop integrals exhibit infrared (IR)/collinear (M)
singularities up to ${\cal O}(\ve^{-2})$ one needs to know the one-loop 
integrals up to ${\cal O}(\ve^2)$ since the one-loop contributions appear 
in product form in the loop-by-loop contributions. 

The aim of our project is thus to compute all one-loop contributions to the two 
processes 
\begin{enumerate}
\item $g+g \rightarrow Q + \bar{Q}$
\item $q+q \rightarrow Q + \bar{Q}$
\end{enumerate}
up to ${\cal O}(\ve^2)$.  

Regarding the four classes of diagrams in Fig.1 one might
then say in a very loose sense that we are aiming to calculate one-fourth of the 
NNLO partonic contributions to heavy hadron production. Nevertheless, calculating 
this one-fourth of the full problem already allows one to obtain a glimpse of the 
complexity that is waiting for us in the full NNLO calculation. This complexity
does in fact reveal itself in terms of a very rich polylogarithmic structure of the 
Laurent series expansion of the scalar one-loop integrals as will be discussed 
later on.

In dimensional regularization there are three different sources that can contribute 
positive $\ve$--powers to the Laurent series of the one--loop amplitudes. These are
\begin{enumerate}
\item Laurent series expansion of scalar one-loop integrals
\item evaluation of the spin algebra of the loop amplitudes bringing in
      the $n$--dimensional metric contraction $g_{\mu \nu}g^{\mu \nu}=n=4-2\ve$
\item Passarino--Veltman decomposition of tensor integrals involving again the
      metric contraction $g_{\mu \nu}g^{\mu \nu}=n=4-2\ve$
\end{enumerate}

Concerning the first item the ${\cal O}(\ve^2)$ calculation of the necessary
one--, two-- and three--point one--loop integrals for the loop--by--loop part of
NNLO QCD calculation have now been completed by us. Two of the three massive one--loop
four-point integrals have also been done leaving us with one remaining four-point
integral which is presently being worked out.
We hope to be able to present complete results on this part of the NNLO calculation 
in the near future \cite{kmr1}.

Concerning the last two items (spin algebra and Passarino--Veltman decomposition)
there exist some partial results on this part of the NNLO calculation which will
be given at the end of this presentation.

Apart from the present discussion of NNLO contributions to heavy hadron
production the calculation of massive scalar loop integrals up to a given 
positive power of $\ve$ is of interest also in other contexts. For example,
if the one--loop integrals appear as subdiagrams in a given divergent
Feynman diagram one again needs to avail of the positive $\ve$--powers 
of the subdiagram. This is of relevance for the calculation of two--loop counter 
terms. Another example is the reduction of a given set of loop 
integrals to master integrals by the integration--by--parts technique 
\cite{ibp}. In the reduction one may encounter
explicit inverse powers of $\ve$ which implies that one has to evaluate
the master integrals up to positive $\ve$-- powers.

\begin{table*}[htb]
\caption{List of scalar n--point functions up to ${\cal O}( \varepsilon^{2})$}
\label{table:5}
\newcommand{\m}{\hphantom{$-$}}
\newcommand{\cc}[1]{\multicolumn{1}{c}{#1}}

{\small
\begin{tabular}{@{}lclcccl}
    \hline\hline
& &Nomenclature of Beenakker {\it et al.} \cite{been} &Our nomenclature &Novelty && Comments\\  \hline
1-point & &       $A(m)$     &  $A$  & -- && Re $\surd$  \\
\hline
2-point & & $B(p_4-p_2,0,m)$ & $B_1$ & -- && Re  $\surd$ \\
        & & $B(p_3+p_4,m,m)$ & $B_2$ & -- && Re, Im $\surd$  \\
        & & $B(p_4,0,m)$     & $B_3$ & -- && Re $\surd$  \\
        & & $B(p_2,m,m)$     & $B_4$ & -- && Re $\surd$   \\
        & &  $B(p_3+p_4,0,0)$ & $B_5$ & -- && Re, Im $\surd$   \\
\hline
3-point & & $C(p_4,p_3,0,m,0)$ & $C_1$ & new && Re, Im $\surd$  \\
        & & $C(p_4,-p_2,0,m,m)$ & $C_2$ & new && Re $\surd$ \\
        & & $C(-p_2,p_4,0,0,m)$ & $C_3$ & -- && Re $\surd$  \\ 
        & & $C(-p_2,-p_1,0,0,0)$ & $C_4$ & -- && Re, Im $\surd$  \\ 
        & & $C(-p_2,-p_1,m,m,m)$ & $C_5$ & new && Re, Im $\surd$ \\ 
        & & $C(p_3,p_4,m,0,m)$ & $C_6$ & -- && Re, Im $\surd$  \\ 
\hline
4-point & & $D(p_4,-p_2,-p_1,0,m,m,m)$ & $D_1$ & new && Re, Im   \\
        & & $D(-p_2,p_4,p_3,0,0,m,0)$ & $D_2$ & new && Re, Im $\surd$  \\
        & & $D(-p_2,p_4,-p_1,0,0,m,m)$ & $D_3$ & new && Re $\surd$ \\ \hline\hline
\end{tabular}
}
\end{table*} 

\section{Laurent series expansion of scalar one-loop integrals}

In Table 1 we provide a list of the one-loop scalar one--, two--, three-- and 
four--point integrals that need to be evaluated up to ${\cal O}(\ve^2)$ in
NNLO heavy hadron production. In column 2 the integrals are identified using the
notation of \cite{been}. When writing down a Laurent series expansion of these 
integrals in terms of $\ve$--powers one needs to introduce a short--hand notation for
the integrals in order to keep the notation managable. Our short--hand notation 
for the integrals
appears in column 3. In column 5 we comment on their reality property and tick
off those integrals that have been completed up to now. As mentioned before
all one--loop integrals have been done except for the four--point integral
$D_1$ involving three massive propagators. We have compared our results to results 
in the literature whenever possible and when these were accessible. We have
found agreement. In column 4, finally, we indicate which of our results are new.

We mention that one also
needs the imaginary parts of the amplitudes since the
square of the amplitude contains also imaginary parts according to 
\begin{equation}
|A|^2 = (Re A)^2 + (Im A)^2 .
\end{equation}
Note, though, that the imaginary parts are only needed up to ${\cal O}(\ve^1)$ 
since the $IR/M$ singularities in the imaginary parts of the one--loop
contributions are of ${\cal O}(\ve^{-1})$ only.

The scalar four--point integrals are the most difficult to calculate. They contain a
very rich structure in terms of polylogarithmic functions. For example, the 
$\ve^2$--coefficients of the Laurent series expansion of the
four--point integrals contain logarithms and classical 
polylogarithms up to order four ({\it i.e.} $ Li_4 $) (in conjunction with 
the $\zeta$--functions $\zeta(2,3,4)$) and products thereof, and a new class of 
functions which are now termed multiple polylogarithms \cite{gon}.
A multiple polylogarithm is represented by
\begin{eqnarray}
\lefteqn{Li_{m_{k},...,m_{1}}(x_{k},...,x_{1})=
\int \limits_{0}^{x_{1}x_{2}...x_{k}} \left( \frac{dt}{t} \circ \right)^{m_{1}-1}} \nonumber \\ 
& &\frac{dt}{x_{2}x_{3}...x_{k}-t} \circ 
\left( \frac{dt}{t} \circ \right)^{m_{2}-1} 
                                                                     \nonumber \\ 
& &  \frac{dt}{x_{3}...x_{k}-t} 
\circ ...  \circ  \left( \frac{dt}{t} \circ \right)^{m_{k}-1}   \frac{dt}{1-t}\; ,
\nonumber
\end{eqnarray}
where the iterated integrals are defined by
\begin{eqnarray}
\lefteqn{\int \limits_{0}^{\lambda} \frac{dt}{a_{n}-t}\circ ...\circ  
\frac{dt}{a_{1}-t}=} \nonumber \\
& & \!\!\!\!\!\!\int \limits_{0}^{\lambda} \frac{dt_{n}}{a_{n}-t_{n}}
\int \limits_{0}^{t_{n}} \frac{dt_{n}}{a_{n-1}-t_{n-1}} \times...\times \int \limits_{0}^{t_{2}}\frac{dt}{a_{1}-t_{1}} \;.\nonumber
\end{eqnarray}

Instead of using the multiple polylogarithms of Goncharov we have chosen to write 
our results in terms of one-dimensional
integral representations given by the integrals 
\begin{eqnarray}
\lefteqn{F_{\sigma_1\sigma_2\sigma_3}(\alpha_1,\alpha_2,\alpha_3,\alpha_4)=}
                                                                       \nonumber \\
& &\int_0^1 dy \frac{\ln (\alpha_1+\sigma_1 y) \ln (\alpha_2+\sigma_2 y)
\ln (\alpha_3+\sigma_3 y)}{\alpha_4+y} \nonumber
\end{eqnarray}
and
\begin{eqnarray}
\lefteqn{F_{\sigma_1}(\alpha_1,\alpha_2,\alpha_3,\alpha_4)=} \nonumber \\
& &\int_0^1 dy \frac{\ln (\alpha_1+\sigma_1 y) {\rm Li}_2(\alpha_2+\alpha_3 y)
}{\alpha_4+y}\ ,\nonumber
\end{eqnarray}
\noindent where the $\sigma_i $ take values $\pm 1$ and the $\alpha_j$'s are 
combinations of the kinematical variables of the process. The numerical evaluation
of these one-dimensional integral representations are quite stable.

A comment on the length of our expressions is appropiate.
The untreated computer output of the integrations is generally quite lengthy. The
hard work is to simplify these expressions. We have written semi-automatic computer
codes that achieve the simplifications using known identities among polylogarithms.
All intermediate steps have been checked numerically. Even after simplification, 
the final expressions are generally too long to be given in this presentation. As 
mentioned earlier on, we have compared our results to the results of other authors 
when they were available and have found agreement.

\section{Laurent series expansion of one--loop amplitudes}
  
In order to obtain the Laurent series expansion of the full one--loop amplitude
one has to combine the Laurent series expansion of the 
scalar one-loop integrals with the $\ve$--expansion from the spin algebra
calculation and the Passarino--Veltman decomposition. This is the subject of a
second paper which we are working on \cite{kmr2}. As an example of such an amplitude 
result we present results on the contribution of the 
one--loop gluon triangle graph to the process $g+g \rightarrow Q + \bar{Q}$.
We shall only list the $\varepsilon$-- and $\varepsilon^2$--terms for the 
triangle graph contribution since the finite and divergent pieces have been 
calculated before (see e.g. \cite{km}). 

Let us begin by defining a reduced
amplitude which is obtained by extracting the spin wave functions from the full
amplitude M, {\it i.e.}
\begin{equation} 
M=\ve(p_1)\ve(p_2)\bar{u}(p_3)M^{\mu \nu}u(p_4).
\end{equation}

For the reduced gluon triangle amplitude $M_{\rm (tri)}^{\mu\nu}(g)$
with gluons and ghosts inside the triangle loop one obtains
\begin{eqnarray}
\label{glutri}
\lefteqn{M_{\rm (tri)}^{\mu\nu}(g) =  N_C \{ -3\,\varepsilon (B_s^{\mu\nu} 
                                 [207 B_5^{(1)} + 12 B_5^{(0)}} \nonumber  \\
&& + 54 C_4^{(1)} s + 8] + 6 i (T^a T^b - T^b T^a) {\rm \hspace{-.1in}}  \not p_1
( g^{\mu\nu}                                                     \nonumber \\
&&[9 B_5^{(1)}\!\!-\! 12 B_5^{(0)}\!\! + 9 C_4^{(1)} s\! - 8]/s 
                            +\! 8 p_2^{\mu} p_1^{\nu}\, [3 B_5^{(0)} \nonumber \\ 
&& + 2]/s^2 )) -  \varepsilon^2 (B_s^{\mu\nu} [
                     621 B_5^{(2)}\!\! + 36 B_5^{(1)}\!\! + 24 B_5^{(0)} \nonumber \\
&& + 162 C_4^{(2)} s + 16]  + 6 i (T^a T^b - T^b T^a) {\rm \hspace{-.1in}}  
                                            \not p_1( g^{\mu\nu} \nonumber \\  
&&[27 B_5^{(2)}\! - 36 B_5^{(1)}\! - 24 B_5^{(0)} + 
                                          27 C_4^{(2)} s - 16]/s \nonumber \\    
&& + 8 p_2^{\mu} p_1^{\nu} [9 B_5^{(1)} + 6 B_5^{(0)} + 4]/s^2 )) \}/324. 
\end{eqnarray}
The one-loop triangle graph amplitude $M_{\rm (tri)}^{\mu\nu}(g)$ has been 
written in terms of the Laurent series expansion of the two massless scalar two--point
and three--point one--loop integrals $B_5$ and $C_5$ listed in Table 1, where 
the Laurent series expansions are defined by 
\begin{equation}
B_5=i C_{\ve}(m^2)\bigg(\sum_{n=-1}^N \ve^n B_5^{(n)}\bigg),
\end{equation}
and
\begin{equation}
C_4=i C_{\ve}(m^2)\bigg(\sum_{n=-2}^N \ve^n C_4^{(n)}\bigg) \, ,
\end{equation}
where $N=2$ in our application.
The Born term amplitude $B_s^{\mu\nu}$ appearing in (\ref{glutri}) is defined by
\cite{km}
\begin{equation}
B_s^{\mu\nu}=2i(T^aT^b\!-T^bT^a)(g^{\mu\nu}{\rm \hspace{-.1in}}  \not p_1
+ p_2^\mu \gamma^\nu\! -p_1^\nu \gamma^\mu)/s ,
\end{equation}
and the coefficient $C_{\ve}(m^2)$ is given by 
\begin{equation}
C_\ve(m^2)= \frac{\Gamma(1+\ve)}{(4\pi)^2}\left(\frac{4\pi \mu^2}{m^2}\right)^2 .
\end{equation}

After insertion of the appropiate coefficient functions $B_5^{(n)}$ and
$C_4^{(n)}$ our results can be seen to fully agree with the results of 
the authors of \cite{davyd} who also calculated the gluonic one--loop corrections 
to the three--gluon vertex with one off-shell gluon. 

Eq.(\ref{glutri}) gives an impression of the interplay of the three different
sources of positive $\ve$--powers in the amplitude calculation mentioned earlier on. 
Note in particular that
different orders of the Laurent series coefficients of the scalar integrals
enter at each order of the Laurent series expansion of the full amplitude.

\vspace{1cm} {\bf Acknowledgements:}
We would like to thank the organizers of the 2004 Loops and Legs Conference
in Zinnowitz, J. Bl\"umlein, S. Moch and T. Riemann, for having provided 
such a nice setting for the conference at the Baltic Sea together with a very 
lively and stimulating scientific athmosphere.

\end{document}